\documentclass{elsart}

\input epsf

\begin{document}

\begin{frontmatter}

\title{Study of the decay mode $D^0 \rightarrow K^-K^-K^+\pi^+$}

\begin{center}  \Large FOCUS Collaboration \end{center}

\author{J.~M.~Link$^1$,}
\author{P.~M.~Yager$^1$,}	    
\author{J.~C.~Anjos$^2$,}
\author{I.~Bediaga$^2$,}
\author{C.~G\"obel$^2$,}
\author{J.~Magnin$^2$,}
\author{A.~Massafferri$^2$,}
\author{J.~M.~de~Miranda$^2$,}
\author{A.~A.Machado,$^2$,}
\author{I.~M.~Pepe$^2$,}
% ...

\begin{abstract}
Using data from the FOCUS (E831) experiment at Fermilab, we present a new 
measurement of the  branching ratio for the Cabibbo-favored decay mode 
$D^0 \rightarrow K^-K^-K^+\pi^+$.

From a sample of $143 \pm 19$ fully reconstructed $D^0 \to K^-K^-K^+\pi^+$ events,
we measure
$\Gamma(D^0 \to K^-K^-K^+\pi^+)/\Gamma(D^0 \to K^-\pi^-\pi^+\pi^+) = 0.00257 \pm 0.00034(stat.) 
\pm 0.00024(syst.)$.
   
 A coherent amplitude analysis has been performed  to determine the resonant 
substructure of this decay mode. This analysis reveals a dominant contribution from
$\phi$ and $\overline K^{*0}(892)$ states.

\end{abstract}

\end{frontmatter}  

\textbf{1. Introduction}

Hadronic decays of charm mesons have been extensively studied in
recent years. Dalitz plot analyses of $D$ meson decays in three-body final states have revealed a 
rich resonant substructure, showing a dominance of quasi two-body modes. However, much less
information is available on the resonant substructure of decays with more than three final state 
particles. These  multi-body $D$ decays account for a large fraction of the hadronic decay width.
If the amplitude analyses of multi-body decays confirm the picture drawn by the Dalitz plot analyses,
then one can make a more complete comparison with theoretical models, which have been developed 
mainly to describe the two-body and quasi-two-body decay modes~\cite{BSW,BLOCK,BLMP,CHAU,BEDAQUE}.  

 We present a new study of the $D^0 \to K^-K^-K^+\pi^+$ decay using data from the 
FOCUS experiment. This is an interesting decay mode: although Cabibbo favored, it is 
strongly suppressed by phase-space and requires the production of an 
$s\overline{s}$ pair, either from the vacuum or via final state interactions (FSI).
We measure the  branching ratio
${\frac{{\Gamma(D^0 \to K^-K^-K^+\pi^+)} }{{\Gamma(D^{0} \to K^-\pi^-\pi^+\pi^+)}}}$ and perform
a coherent amplitude analysis to determine its resonant substructure.

 FOCUS is a charm photoproduction experiment~\cite{spectro} which collected 
data during the 1996--97 fixed-target run at Fermilab. Electron and positron beams (with
typically $300~\textrm{GeV}$ endpoint energy) obtained from the $800~\textrm{GeV/$c$}$ Tevatron
proton beam produce, by means of bremsstrahlung, a photon beam which
interacts with a segmented BeO target. The mean photon energy for triggered
events is $\sim 180~\textrm{GeV}$. A system of three multicell threshold \v{C}erenkov
counters performs the charged particle identification, separating kaons from
pions with momenta up to $60~\textrm{GeV}/c$. Two systems of silicon microvertex
detectors are used to track particles: the first system consists of 4 planes
of microstrips interleaved with the experimental target~\cite{WJohns}; the
second system consists of 12 planes of microstrips located downstream of the
target. These detectors provide high resolution in the transverse plane
(approximately $9~\mu\textrm{m}$), allowing the identification and separation of charm
primary (production) and secondary (decay) vertices. The charged particle
momenta are determined by measuring their deflections in two magnets of
opposite polarity through five stations of multiwire proportional chambers.

%

\vskip 0.5cm \textbf{2. Analysis of the decay mode $D^0 \to K^-K^-K^+\pi^+$} %

The final states are selected using a \textit{candidate driven vertex
algorithm}~\cite{spectro}. A secondary vertex is formed from the four
candidate tracks. The momentum of the resultant $D^{0}$ candidate is used as
a \textit{seed} track to intersect the other reconstructed tracks to 
search for a primary vertex. The confidence levels of both vertices are
required to be greater than $1\%$. 
Once the production and decay vertices are determined, the distance $\ell$ between 
them and its error $\sigma_{\ell}$ are computed. The quantity 
$\ell / \sigma_{\ell}$ is an unbiased measure of the significance of 
detachment between the primary and secondary vertices. 
This is the most important variable for separating charm events from non-charm, prompt
backgrounds. Signal quality is further enhanced by cutting on \emph{Iso2}, which is the confidence
level that other tracks in the event might be associated with the secondary vertex.
To minimize systematic errors on the measurements of 
the branching ratio, we use identical vertex cuts on the signal and normalizing mode, 
namely  
$\ell /\sigma_{\ell}$ $>$ $6$, 
and \emph{Iso2}~$<$~1~$\%$. We also require the $D^{0}$
momentum to be in the range 25--250~\textrm{GeV}/$c$ (a very loose cut) and
the primary vertex to be formed with at least two reconstructed tracks
in addition to the $D^0$ seed. 

The only difference in the selection criteria between the two decay modes 
lies in the particle identification cuts.
The \v{C}erenkov identification cuts used in
FOCUS are based on likelihood ratios between the various particle
identification hypotheses. These likelihoods are computed for a given track
from the observed firing response (on or off) of all cells
within the track's ($\beta =1$) \v{C}erenkov cone for each of our three 
\v{C}erenkov counters. The product of all firing probabilities for all the cells
within the three \v{C}erenkov cones produces a $\chi ^{2}$-like variable 
$W_{i}=-2\ln (\mathrm{Likelihood})$ where $i$ ranges over the electron, pion,
kaon and proton hypotheses~\cite{cerenkov}. All kaon tracks are required
to have $\Delta _{K}=W_{\pi }-W_{K}$ (kaonicity) greater than $2$, whereas 
the pion tracks are required to have $\Delta _{\pi }=W_{K}-W_{\pi }$ (pionicity)
exceeding $0.5$. 

Using the set of selection cuts just described, we obtain the
invariant mass distributions for $K^{-}K^{-}K^{+}\pi^{-}$ and $K^{-}\pi^{-}\pi^{+}\pi^{+}$
shown in Fig.~\ref{4bodies}.
In Fig.~\ref{4bodies}a the $K^{-}K^{-}K^{+}\pi^{-}$ mass plot is fit
with two Gaussians with the same mean but different sigmas to take into 
account the change in resolution with momentum of our spectrometer~\cite{spectro} 
plus a second-order polynomial. A log-likelihood fit returns a signal of $143 \pm 19$
$D^0 \to K^{-}K^{-}K^{+}\pi^{-}$ events.
The large statistics $K^{-}\pi^{-}\pi^{+}\pi^{+}$ mass plot of Fig.~\ref{4bodies}b is fitted in
the same way  (two Gaussians plus a second-order polynomial). The fit gives a signal 
of $64576\pm 360$  $D^0 \to K^{-}\pi^{-}\pi^{+}\pi^{+}$ events. 

%
%

\vskip 0.5cm \textbf{3. Relative Branching Ratio}

The evaluation of relative branching ratios requires yields from the fits to
be corrected for detection efficiencies, which differ among the various
decay modes due to differences in both spectrometer acceptance (due to
different $Q$ values and resonant substructure of the two decay modes) and \v{C}erenkov
identification efficiency.

From Monte Carlo simulations, we compute the relative efficiency to 
be: 
${\frac{\epsilon(D^0 \to K^-K^-K^+\pi^+) }{\epsilon(D^{0} \to K^-\pi^-\pi^+\pi^+)}}
= 0.862 \pm 0.010$.  Combined with the yields measured above we find a branching ratio of
$\frac{\Gamma(D^0 \to K^-K^-K^+\pi^+)}{\Gamma(D^0 \to K^-\pi^-\pi^+\pi^+)} = 0.00257 \pm 0.00034$.

 Our final measurement has been tested by modifying each of the vertex
and \v{C}erenkov cuts individually. The branching ratio is stable
versus several sets of cuts as shown in Fig.~\ref{Brvscuts} (we vary the confidence
level of the secondary vertex from $1\%$ to $10\%$, \emph{Iso2} from $10^{-6}$ to $1$,
$\ell / \sigma_{\ell}$ from $5$ to $15$, $\Delta _{K}$ from $0.5$ to
$4.5$ and finally $\Delta _{\pi }$ from $0.5$ to $4.5$).

Systematic uncertainties on branching ratio measurements can come from
different sources. We determine four independent contributions to the
systematic uncertainty: the \emph{split sample} component, the \emph{fit
variant} component, the component due to the particular choice of the
vertex and \v{C}erenkov cuts (discussed previously), and the limited statistics 
of the Monte Carlo.

The \emph{split sample} component takes into account the systematics
introduced by a residual difference between data and Monte Carlo, due to either a
possible mismatch in the reproduction of the $D^{0}$ momentum or the change in the
experimental conditions of the spectrometer during data collection. This 
component has been determined by splitting data
into four independent sub-samples, according to the $D^{0}$ momentum range
(high and low momentum) and the configuration of the vertex detector,
that is, before and after the insertion of an upstream silicon system\cite{WJohns}. A technique
employed in FOCUS and in the predecessor experiment E687, modeled after the 
\emph{S-factor method} from the Particle Data Group~\cite{PDG}, is used 
to try to separate true systematic variations from statistical 
fluctuations. The branching ratio is evaluated for each of the 4 
statistically independent sub-samples and a \emph{scaled variance} is calculated $\tilde{\sigma}$ 
(the errors are boosted when $\chi ^{2}/(N-1)>1$).  
The \emph{split sample} variance $\sigma_{split}$ is defined as the 
difference between the reported statistical variance and the scaled variance, 
if the scaled variance exceeds the statistical variance~\cite{brkkpipi}.

Another possible source of systematic uncertainty is the \emph{fit variant}.
This component is computed by varying, in a reasonable manner, the fitting
conditions on the whole data set. In our study we fixed the widths of the 
Gaussians to the values obtained by the Monte Carlo simulation, changed 
the background parametrization (varying the degree of the polynomial), and  
used one Gaussian instead of two. In addition we considered the variation of the
computed efficiency,  both for $D^0 \to K^{-}K^{-}K^{+}\pi^{-}$ and the normalizing decay mode, 
due to the different resonant substructure simulated in the Monte Carlo. 
The BR values obtained by these variants are all a priori likely; therefore this uncertainty 
can be estimated by the {\it rms} of the measurements \cite{brkkpipi}.

Analogous to the \emph{fit variant}, the cut component is estimated using 
the standard deviation of the values obtained from the many sets of cuts shown in 
Fig.~\ref{Brvscuts}.
Actually this is an overestimate of this component because the event samples of 
the various cut sets are different. 

 Finally, there is a contribution due to the limited statistics of 
the Monte Carlo simulation used to determine the efficiencies. 
The resulting systematic errors are summarized in Table~\ref{err_sist}.
Adding in quadrature the four components, we obtain the total systematic 
error also shown in Table~\ref{err_sist}.

\begin{table}[h!]
\begin{center}
\begin{tabular}{|l|c|}
\hline
{Source}        & {Percent}      \\
\hline
{Split sample}  & $ 0.0 \%$      \\
{Fit Variant}   & $ 4.1 \%$      \\
{Set of cuts}   & $ 8.3 \%$      \\
{MC statistics} & $ 1.2 \%$      \\ \hline
{Total systematic } & $ 9.4 \%$  \\ \hline
\end{tabular}
\caption{Contribution in percent to the systematic uncertainties of 
the branching ratio $\Gamma(D^0 \to K^-K^-K^+\pi^+)/\Gamma(D^0 \to K^-\pi^-\pi^+\pi^+)$.}
\label{err_sist}
\end{center}
\end{table}

 The final branching ratio result is shown in Table~\ref{comparison} along with a
comparison to previous measurements.

\begin{table}[h!]
\begin{center}
\begin{tabular}{|l|l|l|}
\hline
Experiment & $\frac{\Gamma(D^0 \to K^-K^-K^+\pi^+)}{\Gamma(D^0 \to K^-\pi^-\pi^+\pi^+)}$ &
Events \\ \hline
E687~\cite{E687}   & $0.0028  \pm 0.0007  \pm 0.0001$ & $ 20 \pm 5 $     \\ 
E791~\cite{E791}   & $0.0054  \pm 0.0016  \pm 0.0008$ & $  18.4 \pm 5.3$ \\ 
FOCUS (this result) & $0.00257 \pm 0.00034 \pm 0.00024$ & $ 143 \pm 19 $   \\ \hline
\end{tabular}
\caption{Branching ratio measurement and comparison with other experiments.}
\label{comparison}
\end{center}
\end{table}

%
%

\vskip 1.5cm \textbf{4. Amplitude analysis of $D^0 \to K^-K^-K^+\pi^+$}

A fully coherent amplitude analysis was performed to determine the
resonant substructure of the $D^0 \to K^-K^-K^+\pi^+$ decay. While a
large number of intermediate states could lead to the $K^-K^-K^+\pi^+$
final state, phase space limitations restrict the possible contributions. 
A plot of the $K^-K^+$ invariant mass shows a clear 
$\phi$ contribution (Fig. 3, top left plot).
It is also possible, in principle, to have $K^-K^+$ contribution via
$f_0(980)$ and $a_0(980)$. There are large uncertainties in the line shape 
of these resonances and in their coupling to $K^-K^+$.  They do not appear to be required
by the fit and are therefore not included.

Contributions from $\kappa (800) K^-K^+$  and $\overline K^{*0}(1430) K^-K^+$ might also be present.
However, these resonances are broad scalar states, with no characteristic angular 
distribution that could distinguish them from the non-resonant mode,
with the present level of statistics.
 
Modes containing a $\overline K^{*0}(892)$ can also contribute, even though the
nominal ~$\overline K^{*0}(892)$ 
mass is just above the kinematical limit of the $K^-\pi^+$ spectrum. Since $\overline K^{*0}(892)$ is
a narrow vector meson, its  contribution can be distinguished 
by the angular distribution of the decay products, even without a clear mass peak.

We consider four decay amplitudes: non-resonant, 
$\phi \overline K^{*0}(892)$, $\phi K^-\pi^+$ and $\overline K^{*0}(892) K^-K^+$.
In our default model, model A, we use all four amplitudes. We also consider a simpler
model, model B, which does not include the two $\overline K^{*0}(892)$ amplitudes.

The formalism used in this amplitude analysis is a straightforward extension
to four-body decays of the usual Dalitz plot fit technique. The
$D^0$ is a spin zero particle, so the four-body decay kinematics are defined by
five degrees of freedom. 

Individual amplitudes, $ A_k$, for each
resonant mode are constructed as a product of form
factors, relativistic Breit-Wigner functions, and spin amplitudes which account for angular momentum
conservation.
We use the Blatt-Weisskopf damping factors \cite{blatt}, $F_l$, as form factors ($l$ is
the orbital angular momentum of the decay vertex).
For the spin amplitudes we use the Lorentz invariant amplitudes \cite{mkiii},
which depend both on the spin of the resonance(s) and the orbital angular momentum. The
relativistic Breit-Wigner is

\[
BW = {1 \over {m^2 - m_0^2 + im_0\Gamma(m)}},
\]
where

\[
\Gamma(m) = \Gamma_0 
\frac{m_0}{m}\left(\frac{p^*}{p^*_0}\right)^{2s+1}\frac{F_s^2}{F_{s0}^2}.
\]

In the above equations $m$ is the  two-body invariant mass, $m_0$ and $s$ are the resonance 
nominal mass and spin, and $p^*=p^*(m)$ is the breakup momentum at resonance mass $m$.

Since there are two identical kaons, each amplitude $A_k$ is Bose-symmetrized.
The overall signal amplitude is a coherent sum of
the individual amplitudes,  $A = \sum_k c_k A_k$,
assuming a constant complex amplitude for the non-resonant mode. 
The coefficients $c_k$ are complex numbers to be determined by the fit.
The overall signal amplitude is corrected on an event-by-event basis 
for the acceptance, which is nearly constant across the phase space.

In the amplitude analysis we have taken events having a $K^-K^-K^+\pi^+$ invariant mass within 
the interval $M_D \pm 10$ MeV/$c^2$. In this interval there are 139 signal and 65 background 
events. The finite detector resolution causes a smearing of the edges of the five-dimensional
phase space. This effect is accounted for by multiplying the overall signal distribution 
by a Gaussian factor, $g(M)$, where $M$ is the $K^-K^+K^-\pi^+$ mass. The normalized
signal probability distribution is

\[
P_S(\phi) = \frac {1}{N_S} \varepsilon(\phi) \rho(\phi)
g(M) \mid \sum
c_k A_k(\phi) \mid^2,
\]
where $\phi$ represents the coordinates of an event in the five dimensional phase space, 
$\varepsilon(\phi)$ is the acceptance function, and $\rho(\phi)$ is the phase space density.

Two types of background events were considered: random $\phi$'s combined with a
$K^-\pi^+$ pair, and random combinations of $K^-K^-K^+\pi^+$. Inspection of the
side bands of the $K^-K^-K^+\pi^+$ mass spectrum indicate that nearly 30\% of 
the background events are of the former type. We assume the random  
$K^-K^-K^+\pi^+$ combinations to be uniformly distributed in phase space, while
for the $\phi$ background we assume an incoherent sum of Breit-Wigners with no
form factors and no angular distribution.
The overall background distribution, kept fixed in the fit, is a weighted, incoherent 
sum of these two components. The overall background distribution
is also corrected for the acceptance (assumed to be the same as for the signal events) 
on an event-by-event basis, and multiplied by an exponential function $b(M)$,  
to account for the detector resolution. The 
normalized background probability distribution is
 
 \[
P_B(\phi) = \frac{1}{N_B}\varepsilon(\phi) \rho(\phi) b(M) \sum b_k B_k(\phi).
\]

An unbinned maximum likelihood fit was performed, minimizing the quantity
$w\equiv -2 \ln{\mathcal{L}}$. The likelihood function, $\mathcal{L}$, is

\[
\mathcal{L} = \prod_{events} \left[P_S(\phi^i) + P_B(\phi^i)\right]
\]

Neither the acceptance function nor the phase space density
depend on the fit parameters $c_k$, so the term 
$-2ln\left[ \varepsilon(\phi)\rho(\phi)\right]$ is irrelevant to the minimization.
The acceptance correction is important only for the normalization
integrals $N_S$ and $N_B$.

Decay fractions are obtained from the coefficients $c_k$ determined by the fit,
after integrating the overall signal amplitude over the phase space:

\[
f_k = \frac{\int d\phi\mid c_k A_k \mid^2}{\int d\phi \mid \sum_j c_j A_j \mid^2}.
\]

We fit the data to both models A and B. We find that model B, with only the
non-resonant and $\phi K^-\pi^+$ contributions, does not
provide a good description of the data. The inclusion of the two $\overline K^{*0}(892)$ 
amplitudes results in a much better fit, with an improvement in $\Delta w$ of 138.

Results from the fit with model A are shown in Table 3. The dominant contribution 
comes from  $D^0 \to \phi \overline K^{*0}(892)$. Adding the contributions from 
the two $\overline K^{*0}(892)$ amplitudes, we see that they account for nearly $70\%$ of the 
total decay width. This is somewhat surprising, given the
very small phase space. In Fig.~\ref{proj}  the $K^+K^-$ and  $K^-\pi^+$ 
projections of events used in the amplitude analysis are superimposed on the fit result (top
two plots). The projections from the background model are shown in the shaded histograms.
The remaining plots are two-dimensional projections of the events in the signal region.

We have performed a log-likelihood test to check our ability to distinguish
between models A and B\@. Two ensembles of 10,000 mini-MC samples were generated, one 
simulated according to model A and another  according to model B\@. For each
sample in each ensemble we compute the quantity $\Delta w = 2 \ln{\mathcal{L}_A} - 2 \ln{\mathcal{L}_B}$, where 
$\mathcal{L}_B$ and $\mathcal{L}_A$ are the likelihoods calculated with models B and A, respectively. The
resulting $\Delta w$ distributions are shown in Fig.~\ref{dfcn}. On the right we see
the $\Delta w$ distribution computed from the model A ensemble, and, on the left, 
$\Delta w$ with the model B ensemble. The two distributions are well separated showing that we can
easily distinguish between these two models. Moreover, the value of $\Delta w$ obtained
from the real data (138) is consistent with the distribution from model A, showing that this 
is indeed a better description of the data.

The goodness-of-fit was assessed in two ways. 
We have estimated the confidence level of the fit using a $\chi^2$ test. 
The five invariants used to 
define the kinematics of this decay are the four $K^-K^+$ and $K^-\pi^+$ masses 
squared, plus either the  $K^-K^-$ or $K^+\pi^+$ mass squared. Due to the limited 
statistics we have integrated over the latter invariant and divided the other four 
into two bins, yielding a total of sixteen 
cells. A $\chi^2$ was computed and the estimated confidence level was 35\%. 
The confidence level obtained with model B was 6$\times$10$^{-11}$.

Given the limited statistics
we have also estimated the confidence level using a method which is often less stringent than 
the $\chi^2$. In this second method the confidence level is estimated using the 
distribution of $w = - 2 \ln{\mathcal{L}}$ from 
an ensemble of mini-MC samples generated with the parameters of model A.
This distribution is approximately 
Gaussian. A confidence level can be estimated by the fraction of samples 
in which the value of $w$ exceeds that of the data.  
With this technique we estimate a CL of 86\% for the
fit with model A.  

As in the branching ratio measurement, we consider \emph{split sample} and
\emph{fit variant} systematic uncertainties, using for the former the same sub-samples
described previously.
The dominant contributions from \emph{fit variant} systematic errors come from
variations on the 
signal amplitudes (removing the Blatt-Weisskopf form factors and replacing the
non-resonant amplitude by $\overline K^{*0}(1430) K^+ K^-$), variations in the background 
relative fractions, and variations of the analysis cuts. \emph{Split sample} and \emph{fit variant}
errors are added in quadrature.  Table~\ref{fit} shows the Model A fit results.  The systematic
errors are the second errors in the fractions and phases.

\begin{table}[htb]
\begin{center}
\begin{tabular}{|c|c|c|c|}     \hline
         mode               &   magnitude  & phase(deg.)             & fraction(\%)  
 \\ \hline \hline
 $\phi \overline K^{*0}(892)$    & 1            &     0                   & 48 $\pm$ 6 $\pm$ 1
 \\ \hline
 $\phi K^- \pi^+$           & 0.60 $\pm$ 0.12 & 194 $\pm$ 24 $\pm$ 8 & 18 $\pm$ 6 $\pm$ 4  
 \\ \hline
 $\overline K^{*0}(892) K^+ K^-$ & 0.65 $\pm$ 0.13 & 255 $\pm$ 15 $\pm$ 4 & 20 $\pm$ 7 $\pm$ 2 
 \\ \hline 
     non-resonant           & 0.55 $\pm$ 0.14 & 278 $\pm$ 16 $\pm$ 42& 15 $\pm$ 6 $\pm$ 2      
 \\ \hline
 
\end{tabular}
\protect\caption {Results from the best fit (Model A). The second error on the fractions 
and phases is systematic.}
\label{fit}  
\end{center}
\end{table}

%
%

\vskip 1.5cm \textbf{5. Conclusions}

 Using data from the FOCUS (E831) experiment at Fermilab, we have studied the Cabibbo-favored 
decay mode $D^0 \rightarrow K^-K^-K^+\pi^+$.
 
 A comparison with the two previous determinations of the relative branching ratio 
{$\frac{\Gamma(D^0 \to K^-K^-K^+\pi^+)}{\Gamma(D^0 \to K^-\pi^-\pi^+\pi^+)}$ shows an impressive
improvement in the accuracy of this measurement. 

A coherent amplitude analysis of the $K^-K^-K^+\pi^+$ final state was performed for the first 
time, showing that the dominant contribution comes from $D^0 \to \phi \overline K^{*0}(892)$. 
This channel, $D^0$ decaying to two vector mesons,
corresponds to 50\% of the $D^0 \to K^-K^-K^+\pi^+$ decay rate.
The $\overline K^{*0}(892)$ amplitudes,
in spite of the limited phase space, account for nearly 70\% of the total decay width.
Looking at the $K^+K^-$ spectrum we see that over 60\% comes from  
$D^0 \to \phi \overline K^{*0}(892)$ and $D^0 \to \phi K^- \pi^+$.

%
%

\vspace{1.cm}

We wish to acknowledge the assistance of the staffs of Fermi National
Accelerator Laboratory, the INFN of Italy, and the physics departments of
the collaborating institutions. This research was supported in part by the
U.~S. National Science Foundation, the U.~S. Department of Energy, the
Italian Istituto Nazionale di Fisica Nucleare and Ministero della Istruzione
Universit\`a e Ricerca, the Brazilian Conselho Nacional de Desenvolvimento
Cient\'{\i}fico e Tecnol\'ogico, CONACyT-M\'exico, and the Korea Research
Foundation of the Korean Ministry of Education.

%
%

%
%

\newpage
\begin{figure}[!!t]
\epsfysize=18.cm \epsfxsize=8.cm \epsfbox{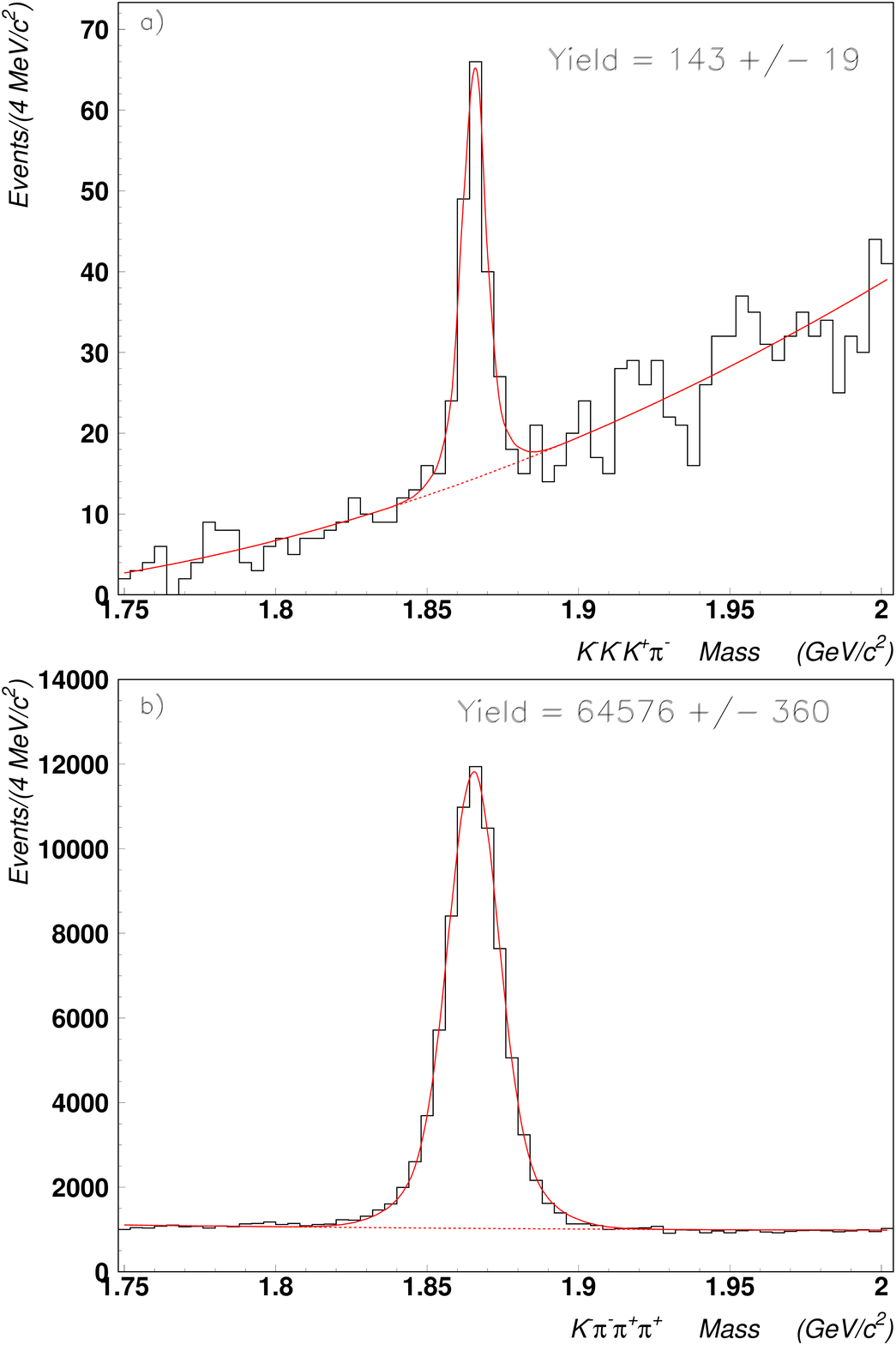} \vspace{0.5cm}
\caption{Invariant mass distribution for $K^-K^-K^+\protect\pi^+$(a)
and $K^-\protect\pi^-\protect\pi^+\protect\pi^+$(b). The fit (solid
curve) is
to two Gaussians for the signal plus a quadratic polynomial for the background.}
\label{4bodies}
\end{figure}
\newpage

\begin{figure}[!!t]
\epsfysize=18.cm \epsfxsize=8.cm \epsfbox{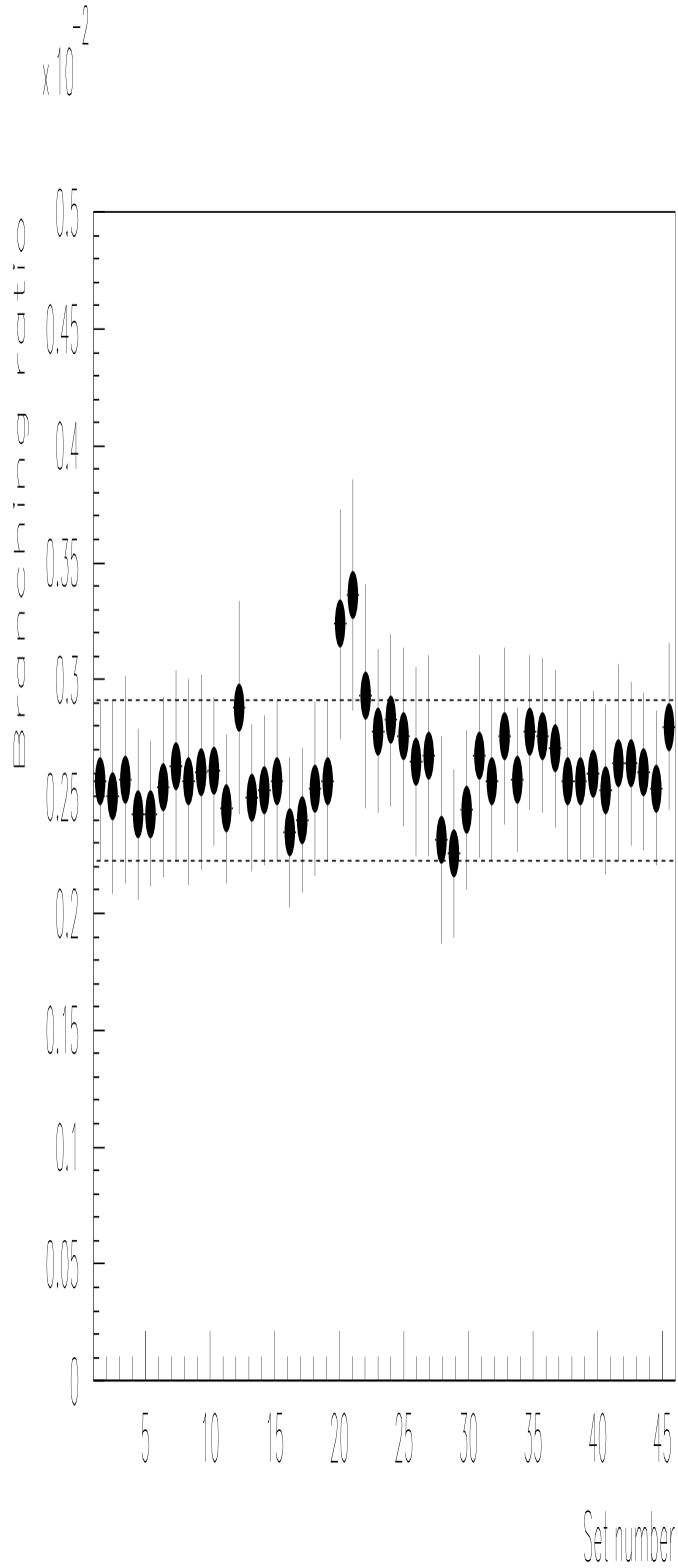} \vspace{0.5cm}
\caption{Branching ratio $\Gamma(D^0 \to K^-K^-K^+\pi^+)/\Gamma(D^0 \to
K^-\pi^-\pi^+\pi^+)$
versus several sets of cuts. From left to right, we vary the confidence
level of the
secondary vertex from  $1\%$ to $10\%$ (10 points), \emph{Iso2} from
$10^{-6}$ to $1$ (7 points), $L$\thinspace  /\thinspace $\sigma _{L}$
from $5$ to $15$ (11 points),
$\Delta _{K}$ from $0.5$ to $4.5$ (9 points) and $\Delta _{\pi }$
from $0.5$ to $4.5$ (9 points).}
\label{Brvscuts}
\end{figure}

\begin{figure}[!!t]
\epsfysize=14.cm \epsfxsize=14.cm \epsfbox{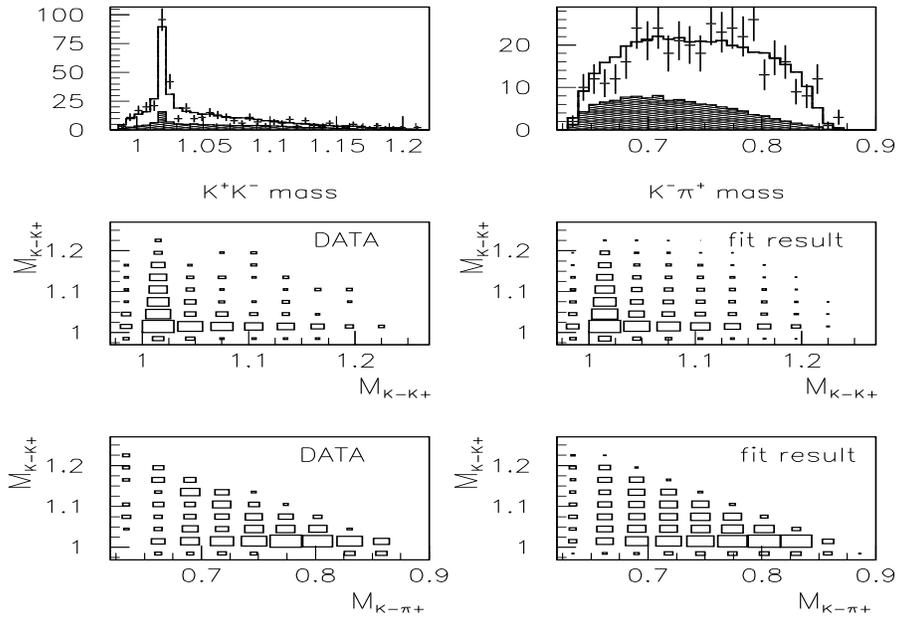} \vspace{0.5cm}
\caption{The top two plots are projections of the invariant mass distribution 
$K^-K^+$ and $K^-\protect\pi^+$, with the fit results superimposed
(solid histograms). The shaded histograms are the background
projections. The left scatter plots show the data population
for $K^-K^+$ and $K^-\protect\pi^+$ from the signal region, while the right ones show the fit results.
Due to two identical particles in the final state, there are two 
entries per event in each plot.}
\label{proj}
\end{figure}

\begin{figure}[!!t]
\epsfysize=18.cm \epsfxsize=14.cm \epsfbox{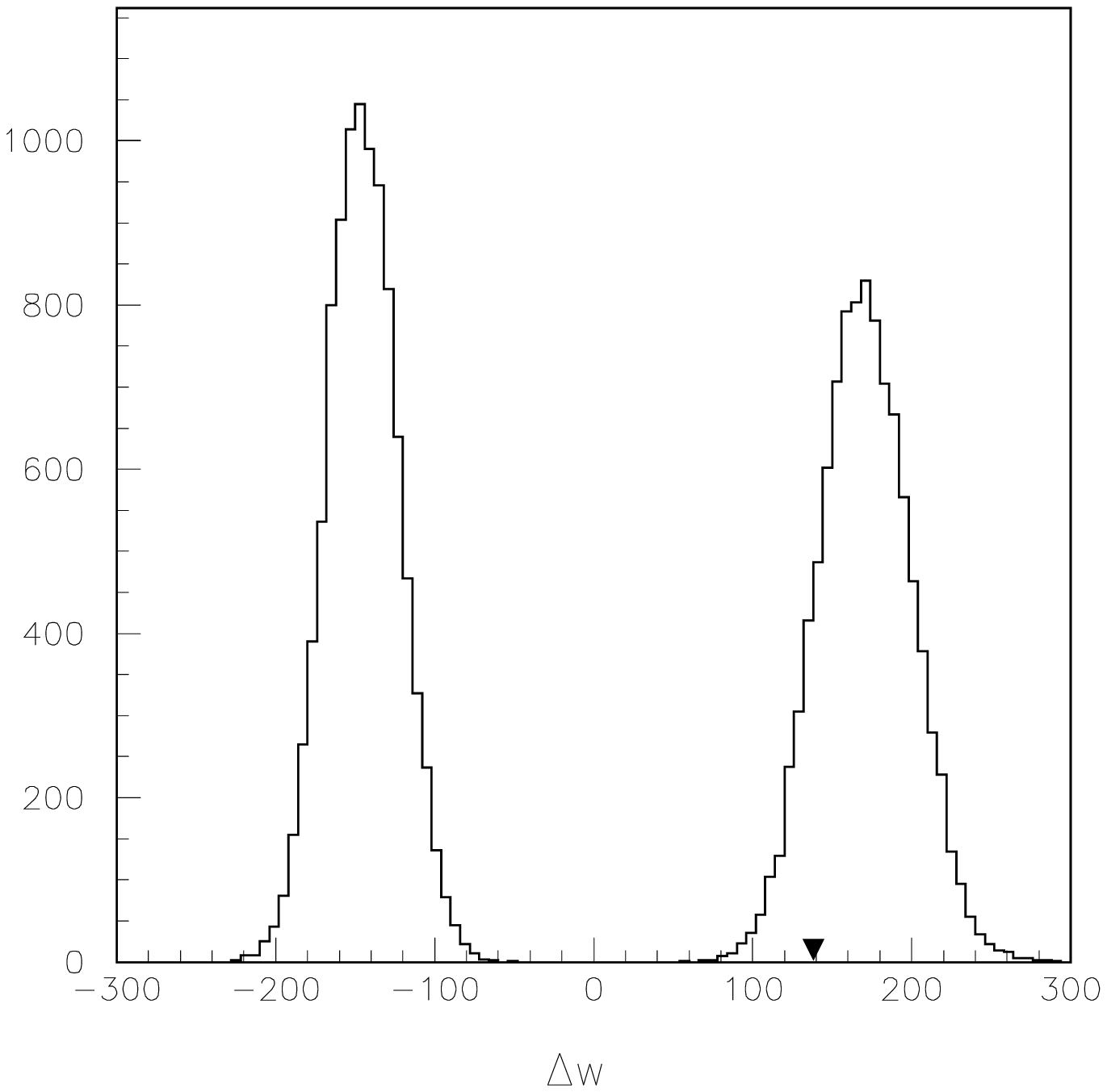} \vspace{0.5cm}
\caption{Distributions of difference in log-likelihood from  ensembles of simulated samples
generated according to model B (left distribution) and model A (right distribution).}
\label{dfcn}
\end{figure}
\newpage


\begin{thebibliography}{99}

\bibitem{BSW} M.~Bauer, B.~Steck and M.~Wirbel, Z. Phys.~C34 (1987) 103.

\bibitem{BLOCK} B.T.~Block and M.A.~Shifman, Sov. J. Nucl. Phys.~45 (1987) 522.

\bibitem{BLMP} F.~Buccella \textit{et al.}, Z. Phys.~C55 (1992) 243.

\bibitem{CHAU} L.L.~Chau and H.Y.~Cheng, Phys. Rev.~D36 (1987) 137.

\bibitem{BEDAQUE} P.~Bedaque, A.~Das, and V.S.~Mathur, Phys. Rev.~D49 (1994) 269.

\bibitem{spectro} E687 Collaboration, P.L.~Frabetti \textit{et al.},
Nucl.~Instrum.~Meth. A320 (1992) 519.

\bibitem{WJohns} FOCUS Collaboration, J.M.~Link \textit{et al.},
\hbox{hep-ex/0204023}, submitted  to Nucl.~Instrum.~Meth. A.

\bibitem{cerenkov} FOCUS Collaboration, J.M.~Link \textit{et al.},
Nucl.~Instrum.~Meth. A484 (2002) 270.

\bibitem{PDG} K.~Hagiwara \textit{et al.} (Particle Data Group), Phys. Rev.~D66 (2002) 010001.


\bibitem{brkkpipi} FOCUS Collaboration, J.M.~Link \textit{et al.}, Phys. Lett.~B555 (2003) 173.

\bibitem{E687} E687 Collaboration, P.L.~Frabetti \textit{et al.}, Phys. Lett.~B354 (1995) 486.

\bibitem{E791} E791 Collaboration, E.M.~Aitala \textit{et al.}, Phys. Rev.~D64 (2001) 112003.

\bibitem{mkiii} MARKIII Collaboration, D.~Coffman {\em{et al.}}, Phys. Rev.~D45 (1992) 2196.

\bibitem{blatt} J.M.~Blatt and V.F.~Weisskopf, {\em Theoretical Nuclear 
Physics}, page 361, John Wiley \& Sons, New York, 1952.

\end{thebibliography}
\end{document}